\documentclass[twocolumn,showpacs,prl,aps,amsmath,amssymb]{revtex4}
\usepackage{graphicx}

\begin{document}

\title{ Electron-phonon interaction in Fe-based superconductors: Coupling of magnetic moments with phonons in LaFeAsO$_{1-x}$F$_{x}$}

\author{Felix Yndurain}
\email{felix.yndurain@uam.es.}
\affiliation{Departamento de F\'{i}sica de la Materia Condensada
and Instituto de Ciencia de Materiales "Nicol\'{a}s Cabrera".
Universidad Aut\'{o}noma de Madrid. Cantoblanco. 28049 Madrid.
Spain.}

\date{\today}

\begin{abstract}
The coupling of Fe magnetic moments in LaFeAsO$_{1-x}$F$_{x}$  with the As $A_{1g}$ phonon is calculated. We present first principles calculations of the atomic and  electronic structure of  LaFeAsO as a function of electron doping. We perform calculations using the virtual crystal approximation as well as supercell calculations with F substitutional impurity atoms. The results validate the virtual crystal approximation for the electronic structure near the Fermi level. Its is found that the electronic density of states at the Fermi level is maximum for x=0.125, enhancing the electron-phonon interaction. An additional increase of the electron-phonon parameter $\lambda$ is obtained if the coupling between the $A_{1g}$ phonon and the Fe magnetic moment is included. It is found that the electron-phonon interaction can be one order of magnitude larger than its value if  no spin resolution is included in the calculation. The implications of these results on the superconducting transition are discussed
\end{abstract}


\maketitle

Since the discovery of the Fe-based superconductors a few years ago \cite{Kamihara-2,Kamihara-1}, several properties common to most of the compounds have been well established and general agreement has been reached. However, there are very important aspects that have not been settled \cite{Review}. The most important one, is the mechanism responsible for their superconducting behavior. Early theoretical work  \cite {Mazin, Boeri} ruled out any phonon's role. More recent calculations \cite{Boeri_Mag} reveal that, in the non-magnetic phase, phonons coupled with magnetism contribute to superconductivity although not enough to explain the high critical temperature. Another aspect to be clarified involves the Fe magnetic moment. The low temperature stripped antiferromagnetic order in the parent compound is generally accepted, although the magnitude of the calculated magnetic moment is much larger than the observed one \cite{Korotin}. Fluctuations mechanisms as well as orbital magnetic ordering \cite{Bascones, OrbitalOrdering} have been proposed to explain this discrepancy. 
The coupling of the Fe magnetic moments with phonons has been clearly established at least in CaFe$_{2}$As$_{2}$, in the undoped orthorombic phase \cite{Phonons_Magnetism}, and in SmFeAsO$_{1-x}$F$_{x}$ \cite{LeTacon} . Also the possibility of a non-zero Fe magnetic moment in the superconducting phase has not been ruled out \cite{Reznik}.

\begin{figure}[ht]
\includegraphics[width=80mm]{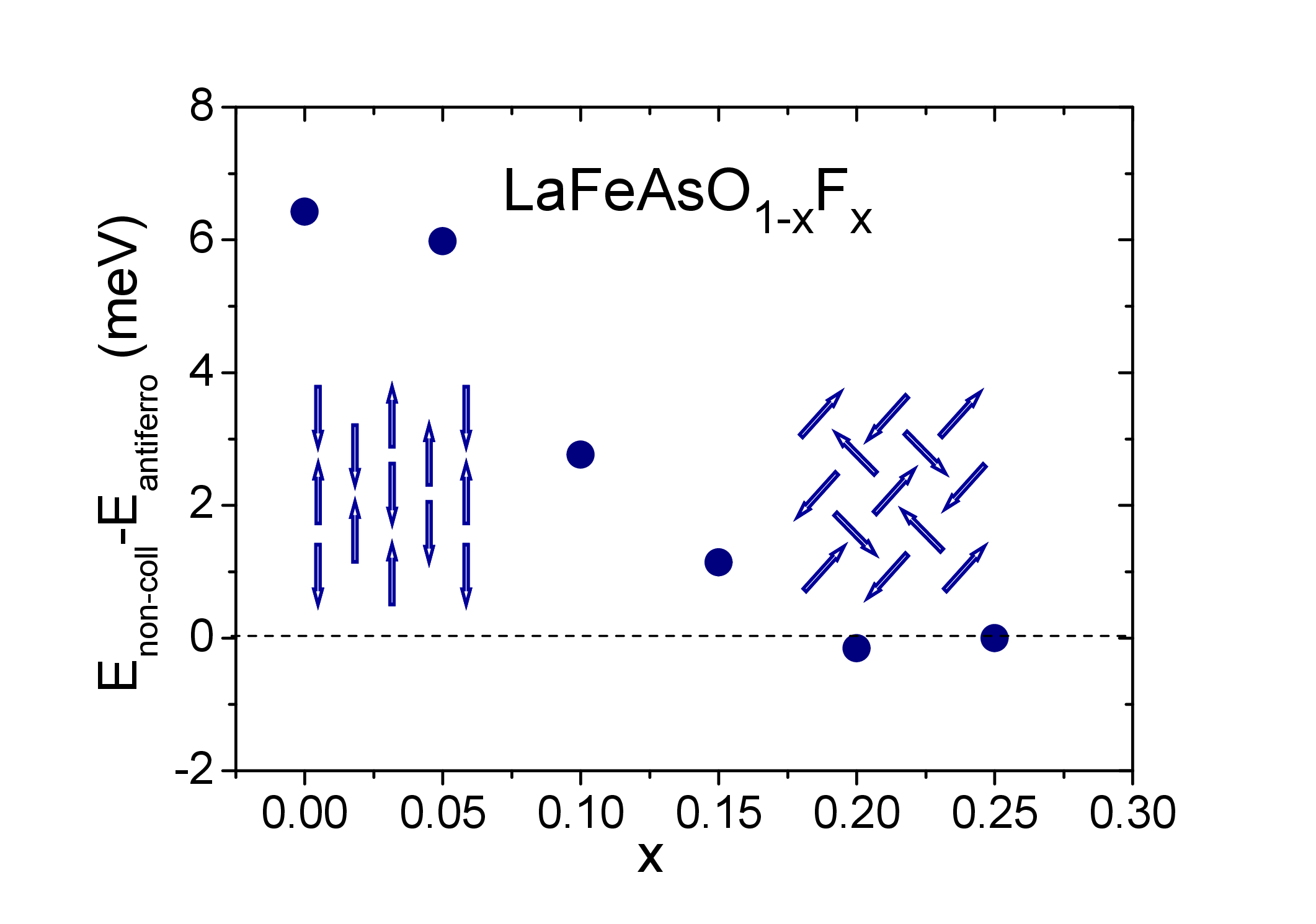}
\caption{(Color online) Energy barrier for the transition between two equivalent antiferromagnetic stripped collinear arrangements (left inset) through an intermediate non collinear state (right inset) in LaFeAsO$_{1-x}$F$_{x}$ as a function of x. The energy difference per formula unit is shown.} \label{figure1}
\end{figure}

Here we propose that the magnetic moment in Fe atoms does not disappear in the superconducting phase. Instead, what disappears is the long-range collinear antiferromagnetic order while keeping the As-mediated antiferromagnetic local coupling between the Fe atoms. To this end, we have calculated in the manner described in \cite{Nosotros} how the energy barrier between two equivalent collinear antiferromagnetic configurations varies with the extra electron concentration in LaFeAsO$_{1-x}$F$_{x}$. We assume that a non-collinear configuration takes place between the two minima  as discussed in \cite{Nosotros}. The results of the calculation are shown in Figure \ref{figure1}. The results indicate that the barrier between two equivalent antiferromagnetic arrangement is very small beyond x=0.1 up to the disappearance of the magnetic moment  around x=0.3. This low barrier can be responsible for the  disappearance of the antiferromagnetic long-range order since the system may fluctuate between the two equivalent magnetic arrangements.

To study how the electronic structure, and eventually, the electron-phonon interaction varies with doping we restrict ourself to the antiferromagnetic collinear arrangement since the non-collinear one has a similar electronic structure (see \cite{Nosotros}) and its calculation is much more computational demanding. To obtain the electronic structure we have performed  density functional \cite {DFT1, DFT2} calculations using the SIESTA code \cite {Siesta1, Siesta2} which uses localized orbitals as basis functions \cite{Orbitals}. In our calculation we use a double $\zeta$ polarized basis set, non-local norm conserving pseudopotentials and a local density approximation (LDA) for  exchange and correlation. The calculations are performed with stringent criteria in the electronic structure convergence (down to $10^{-5}$ in the density matrix), Brillouin zone  sampling (up to 18000 $k$-points), real space grid (energy cut-off of 500 Ryd) and equilibrium geometry (residual forces lower than $10^{-2}$ eV/\AA). Due to the rapid variation of the density of states at the Fermi level, we used a polynomial smearing method \cite{smearing}. To simulate the effect of doping we use the virtual crystal approximation (VCA) \cite{VCA}.

\begin{figure}[h]
\includegraphics[width=80mm]{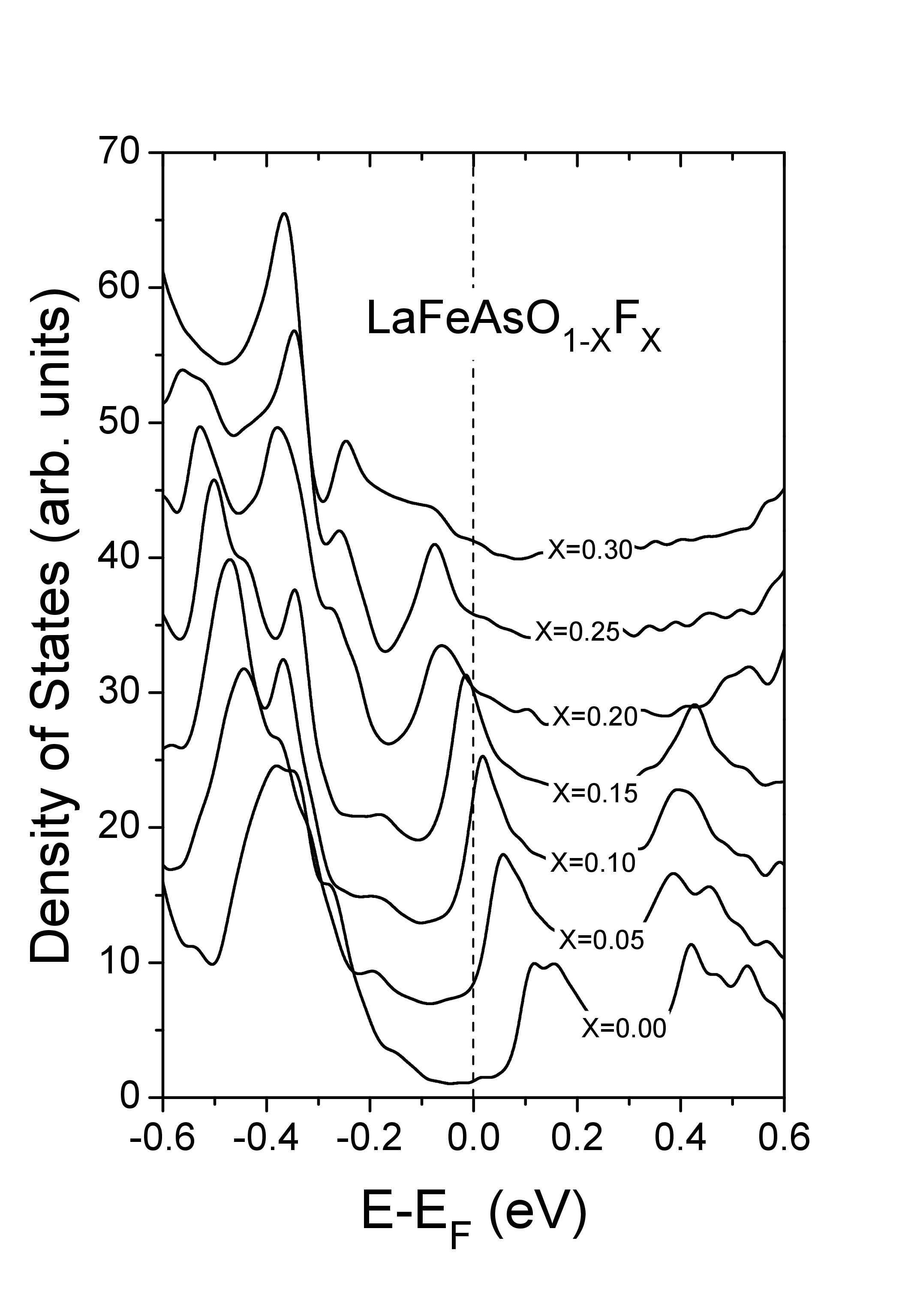}
\caption{(Color online) Variation of the total density of states of LaFeAsO$_{1-x}$F$_{x}$  with the F content x. The calculation is for the virtual crystal approximation. A small imaginary contribution has been added to the energy to smooth the curves.} \label{DOS-versus-x}
\end{figure}

The calculated total densities of states for various fluorine contents are drawn in Figure \ref{DOS-versus-x}. We first observe the antiferromagnetic pseudo gap at the Fermi level for x=0. As the excess of electrons increases with the fluorine concentration the size of the gap decreases and eventually disappears  between x=0.25 and x=0.30. It is interesting to notice that the antiferromagnetic gap lies fully bellow the Fermi level for $x\geq 0.15$. Moreover,  the peak in the density of states at the antiferromagnetic gap upper edge crosses the Fermi level between x=0.10 and x=0.15. 

\begin{figure}[ht]
\includegraphics[width=80mm]{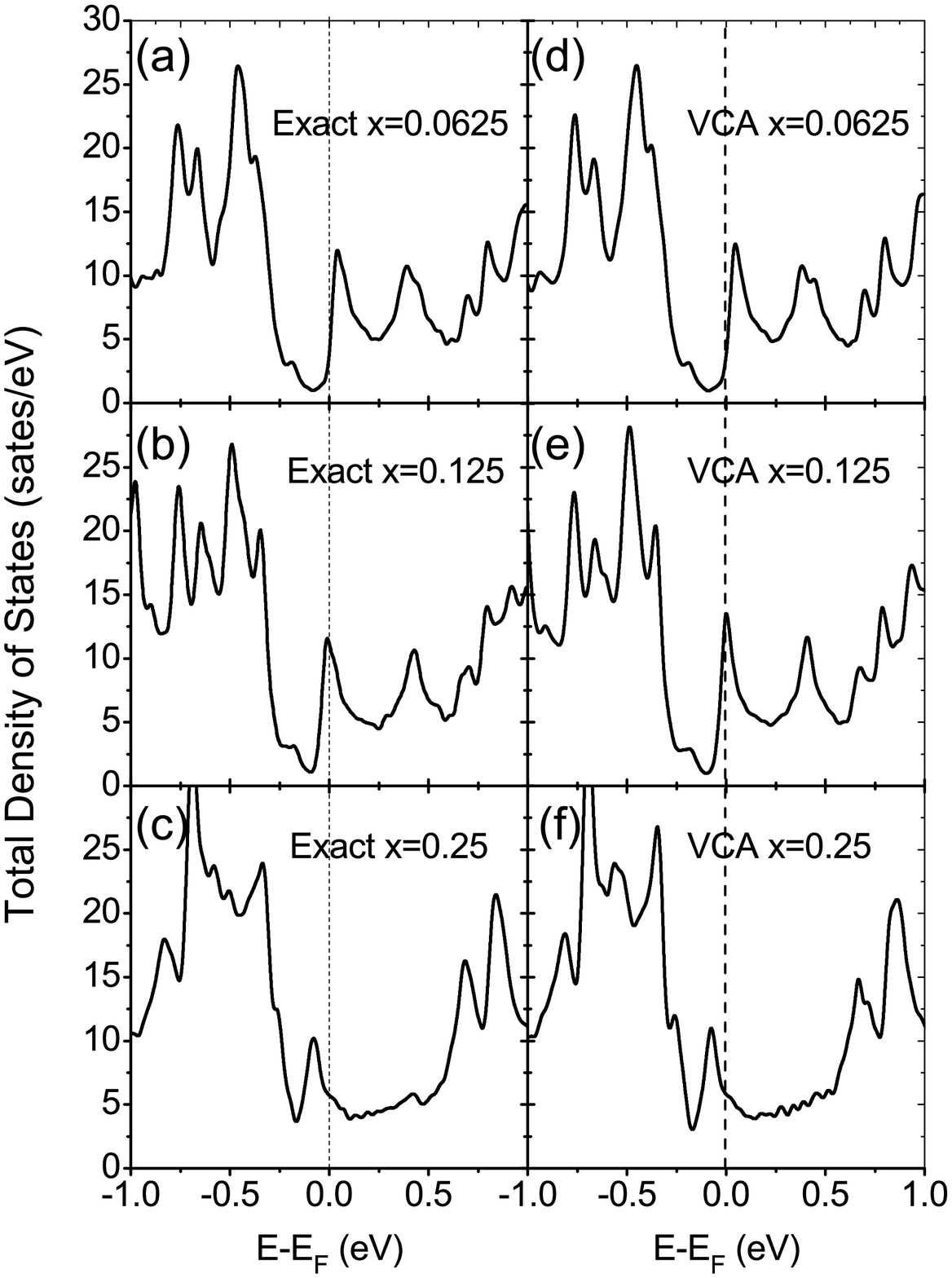}
\caption{Calculated LaFeAsO$_{1-x}$F$_{x}$ electronic densities of states using super-cells ("exact")  (panels (a), (b) and (c)) and using the virtual crystal approximation (panels (d),(e) and (f)) for various values of x. A small imaginary contribution has been added to the energy to smooth the curves.} \label{DOS-Exact-VCA}
\end{figure}

Before we go into the discussion of the electron-phonon interaction parameter we analyze to what extent the virtual crystal approximation is a good starting point to study the electronic structure of these materials. To this end, we have calculated the equilibrium atomic structure and the electronic density of states for various super-cells size substituting an oxygen atom by a fluorine one. In this manner we have performed "exact" calculations for x=0.25, 0.125 and 0.0625. The calculated density of states after full relaxation of the atomic positions are shown in Figure \ref{DOS-Exact-VCA} which shows an excellent agreement with the VCA results. The VCA density of states reproduces the main features of the super-cell calculations, in particular, the peak in the density of sates crossing the Fermi level at around x=0.125. From these results it is clear that the details of how impurities are incorporated in the oxygen layer does not affect much the electronic structure around the Fermi level which is mainly dominated by the Fe electrons. The main effect of doping impurities at the oxygen layer is to pump electrons into the As-Fe-As layer, as anticipated \cite {Takahashi-Nature} .

We have also calculated the influence of an F impurity on the electronic charges and magnetic moments of the Fe atoms. In Table \ref{Table-I} we show the results for a supercell Fe$_{16}$As$_{16}$La$_{16}$O$_{15}$F$_{1}$ which corresponds to x=0.0625. We notice that the extra charge is distributed rather evenly through the Fe atoms and does not distinguish between those closer to the impurity and the other Fe atoms. On the other hand, although the magnetic moment of the  Fe atoms closer to the F impurity display a magnetic moment larger than the others, the difference is rather small and we can rule out any relevant polaron formation. The antiferromagnetic coupling between successive Fe layer agrees with the experimental magnetic order of this material. The above results stress the appropriateness of the VCA approximation. Also,  there is no significant influence of the presence of F impurities in the geometry of  the Fe-As layers next to it. We find that the Fe-As distance depends on the extra charge at the As-Fe-As layers, rather than on the details concerning the oxygen layer and the substitutional impurities.  

\begin{table}[h]
\caption { Electronic charge and magnetic moment (in Bohr magnetons) at the Fe atoms of Fe$_{16}$As$_{16}$La$_{16}$O$_{15}$F$_{1}$ in the super-cell calculation. The corresponding VCA results are indicated.}
\label{Table-I}
\begin{tabular}{ccccc}
\hline\
Fe Layer &Q$_{1}$ &Q$_{2}$&Q$_{3}$&Q$_{4}$  \\
\hline
1&8.298  & 8.305&8.312 &8.314  \\
2&8.298 &8.306 &8.300 &8.304$^{*}$  \\
3&8.298 & 8.305&8.300  & 8.304$^{*}$  \\
4&8.298  & 8.306& 8.312  &8.313  \\
\hline
\hline
Fe Layer &$\mu_{1}$ & $\mu_{2}$  & $\mu_{3}$ &$\mu_{4}$  \\
\hline
1&  -1.222 &  -1.219  & 1.174& 1.212  \\
2& 1.220& 1.218  & -1.238 & -1.234$^{*}$ \\
3& -1.222& -1.219  & 1.240 & 1.236$^{*}$  \\
4& 1.222 &1.220   & -1.176& -1.213 \\
\hline
\hline
\multicolumn{5}{c}{Q$_{VCA}$: 8.301;  $\mu_{VCA}$: 1.247} \\
\multicolumn{5}{l}{ * Fe atoms closest to the F impurity}\\
\hline
\end{tabular}
\end{table}



\begin{figure}[ht]
\includegraphics[width=80mm]{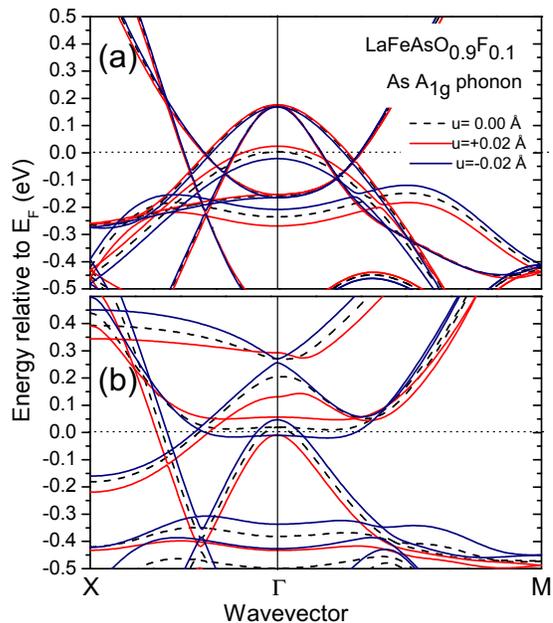}
\caption{(Color online) Band structure with frozen  $A_{1g}$ phonon (vibration of As atoms perpendicularly to the Fe plane) for a doping of 0.10 electrons per formula unit. The results are for the virtual crystal approximation. The dashed bands correspond to zero phonon amplitude and the red (blue) bands correspond to an expansion (compression)  of the Fa-As planes distance $u$ of 0.02 \AA. (a) Spin independent calculation. (b) Stripped antiferromagnetic magnetic order. } \label{BANDS-Para-AF}
\end{figure}

To address the electron-phonon interaction in this system and the superconducting $\lambda$ parameter, we have considered the symmetric out-of-plane As $A_{1g}$ mode at $k_{||}=0$, in which the Fe atoms remain fixed whereas the As atoms move perpendicularly to the FeAs layers, expanding and compressing the Fe-As bonds. For this phonon mode the diagonal electron-phonon matrix element can be written \cite{Allen}:

$ <\overrightarrow k,n\left | H_{e-ph}(\nu ) \right |\overrightarrow k,n>=(\hbar/2M_{a}\omega _{\nu })^{1/2}\epsilon _{\nu}^{a}\cdot \overrightarrow{D}(\overrightarrow k,n)$

where  $\epsilon _{\nu}^{a}$ is the polarization vector, $ \omega _{\nu }$ is the frequency of the vibrational mode involved and $\overrightarrow{D}(\overrightarrow k,n)$ is the deformation potential for the state $(\overrightarrow k,n)$
 
 The $\lambda$ parameter associated to a vibrational mode can be written as a sum of the contributions of the electronic bands crossing the Fermi level in the form: 

$\lambda _{A_{1g}}=\frac{1}{2M_{As}\omega_{A_{1g}} ^{2}}\sum_{n}N_{n}(E_{F})D_{n}^{2}$,

where $N_{n}(E_{F})$ is the nth-band density of states at the Fermi level, $\omega_{A_{1g}}$ the phonon frequency,  $M_{As}$ the mass of the arsenic atoms and $D_{n}$ is the variation of the energy band $E_{n}$ at the Fermi energy the phonon with the phonon amplitude:

$D_{n}=\frac{1}{\sqrt{2}}\frac{dE_{n}(E_{F})}{du}$

Before we calculate the $\lambda$ parameter it is worth looking at the deformation potential associated to the $A_{1g} $ phonon. Although a supercell  approximation could be more precise \cite{Noffsinger}, the calculations are performed within the VCA. In Figure \ref{BANDS-Para-AF} we show, for the fluorine concentration of 10\%, how the electronic band structure is perturbed by the presence of the phonon. In the case of a non magnetic calculation (Figure \ref{BANDS-Para-AF} (a)) the effect of the phonon in the bands at the Fermi level is small in all the bands but one ($d_{xy}$). On the contrary, in the case of the calculation for the antiferromagnetic configuration (Figure \ref{BANDS-Para-AF} (b)), all the bands are perturbed by the presence of the phonon, and, in addition, there is a flattening of the bands due to the antiferromagnetic pseudo-gap. The effect of the  $A_{1g} $ phonon is to modulate the size of the Fe magnetic moment with the vibration amplitude and therefore inducing an additional shift to the bands. A similar behavior was previously discussed \cite{Nosotros}. 
\begin{figure}[ht]
\includegraphics[width=80mm]{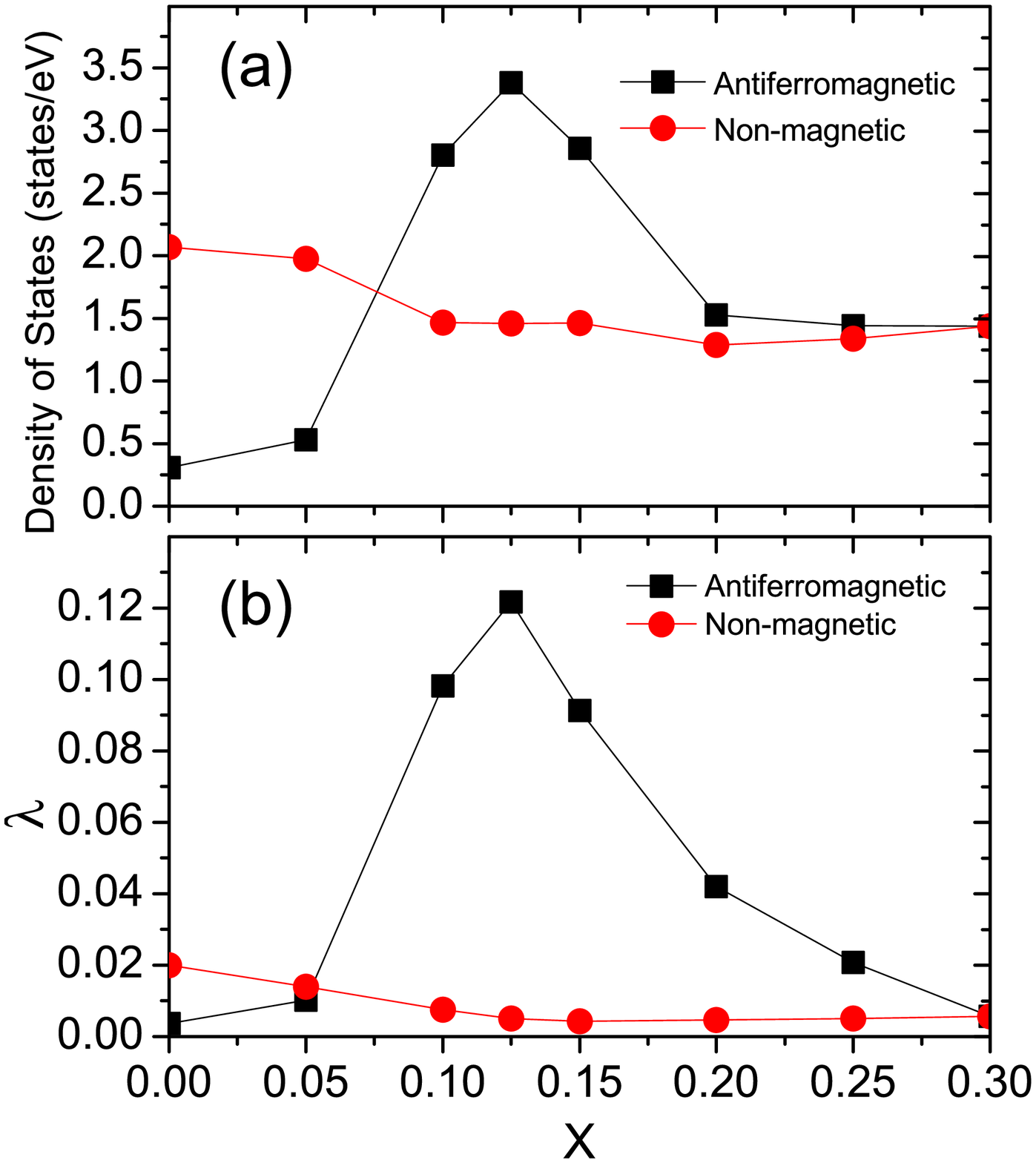}
\caption{(Color on line) Variations of the density of states at the Fermi level and parameter $\lambda$ of LaFeAsO$_{1-x}$F$_{x}$ with the fluorine concentration x. (a) Density of states for the antiferromagnetic collinear order (black squares) and non-magnetic (red circles) calculations. (b) Calculated parameter $\lambda$ for the antiferromagnetic collinear order (black squares) and non-magnetic (red circles)  .} 
\label{Lambda-vs-x}
\end{figure}

We have calculated the $\lambda$ parameter associated to this phonon in the approximation discussed above and within the VCA. The results, along with the variation of the density of states at the Fermi level, are shown in Figure \ref{Lambda-vs-x},
where a phonon energy of 25 meV is considered. We observe that the antiferromagnetic pseudo-gap induces an increase of the density of states at the Fermi energy with respect to the non-magnetic calculation (see Figure \ref{DOS-versus-x}). This increase would enhance the  $\lambda$ parameter by a factor of two at the most. However, if the deformation potential is included, the $\lambda$ parameter can be one order of magnitude larger than its value for the non-magnetic calculation (Figure \ref{Lambda-vs-x} (b)). In addition, if we consider the renormalization of the parameter $\lambda$ due to the renormalization of the phonon mode $\omega ^{2}=\frac{\omega _{0}^{2}}{1+2\lambda }$ then $\lambda =\frac{\lambda _{0}}{1-2 \lambda _{0}}$ which for the maximum value at x=0.125 becomes $\lambda=0.16$. This is indeed a substantial increase with respect to the non-magnetic calculation but not large enough to explain by itself the superconducting critical temperature. All the vibrational modes should be included in the calculation to obtain the total $\lambda$ parameter. The connection between the Fe-As bond length and the magnetic moments in Fe atoms is important (as discussed by Yildirim \cite{yildirim-2}) and is relevant in the electron-phonon interaction.

In summary, we propose that in the superconducting phase of LaFeAsO$_{1-x}$F$_{x}$, the Fe atoms have a finite magnetic moment that fluctuates between two equivalent collinear antiferromagnetic configurations. The presence of this moment varies substantially the density of states near the Fermi level and enhances dramatically the electron-phonon interaction at least fort the As $A_{1g}$ phonon mode.  The maximum of the calculated $\lambda$ parameter takes place at x=0.125. These results suggest that electron-phonon interaction, coupled with the Fe magnetic moments, must be carefully revised, before ruling out its connection with Fe-based compounds superconductivity.

I am indebted to J.M. Soler, for very enlightening discussions and for a critical reading of the manuscript. I also thank  E. Bascones, M. J. Calderon, M.L.Cohen,  G. Gomez-Santos, S.G. Louie, D. Sanchez-Portal and B. Valenzuela  for very helpful comments about this work. This work was supported by the Spanish Ministry of Science and Innovation through grants FIS2009-12712 and CSD2007-00050.

\bibliographystyle{apsrev}

\end{document}